\documentclass[aps,twocolumn,amsmath,amssymb,floatfix,preprintnumbers]{revtex4}

\usepackage{amsmath}
\usepackage{psfig}
\usepackage{fancyhdr}
\usepackage{rotate}
\usepackage{rotating}
\usepackage{dcolumn}
\usepackage{bm}
\usepackage{subfigure}
\usepackage{epsfig}
\usepackage{natbib}
\usepackage{graphicx}
\usepackage{color}
\usepackage{latexsym}
\usepackage{amsmath}

\newcommand{\vect}[1]{\mbox{\boldmath$#1$\unboldmath}}

\newcommand{\ie}{{\it i.e.}}
\newcommand{\eg}{{\it e.g.}}

\newcommand{\grad}{\vect{\nabla}}

\newcommand{\dive}{\vect{\nabla} \cdot}

\begin{document}
  
  
  \title{Reverse-selective diffusion in nanocomposite membranes}
  
  \author{Reghan J. Hill} \email{reghan.hill@mcgill.ca}
  
  \affiliation{Department of Chemical Engineering and McGill Institute
    for Advanced Materials, McGill University\\ Montreal, Quebec,
    CANADA H3A 2B2}
  
  \date{\today}
  
  \begin{abstract}
    The permeability of certain polymer membranes with impenetrable
    nanoinclusions increases with the particle volume
    fraction~(Merkel, {\em et al.}, {\em Science}, 296, 2002). This
    intriguing observation contradicts even qualitative expectations
    based on Maxwell's classical theory of conduction/diffusion in
    composites with homogeneous phases. This letter presents a simple
    theoretical interpretation based on classical models of diffusion
    and polymer physics. An essential feature of the theory is a
    polymer-segment depletion layer at the inclusion-polymer
    interface. The accompanying increase in free volume leads to a
    significant increase in the local penetrant diffusivity, which, in
    turn, increases the bulk permeability while exhibiting reverse
    selectivity. This model captures the observed dependence of the
    bulk permeability on the inclusion size and volume fraction,
    providing a straightforward connection between membrane
    microstructure and performance.
\end{abstract}

\maketitle

Polymer membranes facilitate a variety of molecular
separations. Because their micro-structural scale is comparable to the
size of solute molecules, the polymer architecture can be tailored to
separate specific gas mixtures. Merkel {\em et
al.}~\citep{Merkel:2002} recently showed that incorporating nanometer
sized inorganic particulates into certain amorphous polymer melts
increased the membrane permeability and selectivity. Because of its
significant technological applications, this discovery has stimulated
many experimental
investigations~\citep{Merkel:2003,Gomes:2005,Winberg:2005,Zhong:2005}. A
theory that quantifies how the inclusion size and concentration affect
the permeability and selectivity does not exist. Indeed, Merkel {\em
et al.} highlighted that classical Maxwell-like theories fail to
describe even the qualitative trends.

This letter presents a theoretical interpretation based on classical
models of diffusive transport and polymer physics. Despite its
simplicity, the model accurately describes experimental trends.  The
principal assumption is that there exists a repulsive interaction
between the nanoinclusions and polymer segments. This hypothesis is
support, in part, by transmission electron microscopy (TEM) images of
silica-based nanocomposites~\citep[\eg,][]{Merkel:2002,Bansal:2005},
which show clear evidence of particle aggregation.

For steady gas permeation across a membrane with thickness $L$, the
diffusive flux is
\begin{equation}
  |\langle \vect{j} \rangle| = - D^e \Delta n / L,
\end{equation}
where $D^e$ is the effective diffusivity and $\Delta n$ is the
differential concentration of the diffusing solute. Because the solute
concentrations at the gas-membrane interfaces are proportional to the
partial pressure $p$, with a constant of proportionality $H$ (Henry's
constant), it is customary to express the flux in terms of the
differential partial pressure $\Delta p$:
\begin{equation}
  |\langle \vect{j} \rangle| = -D^e H \Delta p.
\end{equation}
The permeability is defined as $|\langle \vect{j} \rangle| / |\Delta
p| = D^e H$. In this work, the gas solubility is assumed to be
independent of solids concentration, so the inclusions influence the
permeability only through their role in modifying the effective
diffusivity $D^e$.

If voids in the polymer are much larger than the solute molecules,
then the diffusivity in these voids $D$ will be much larger than in
the polymer $D^\infty$. When $D / D^\infty \rightarrow \infty$,
Maxwell's theory yields an effective diffusivity $D^e = D^\infty (1 +
3 \phi_v) + O(\phi_v^2)$, where $\phi_v$ is the (small) volume
fraction of (spherical) voids. Under these conditions, the relative
enhancement of all effective diffusivities are the same, so membrane
selectivity is unaltered.

\begin{figure}
  \begin{center}
    \vspace{0.2cm} \includegraphics[width=5cm]{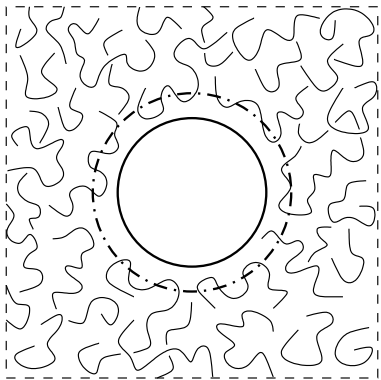}
    \vspace{0.2cm}
    \caption{\label{fig:figure1} An impenetrable nanosphere with
      radius $a \sim 5$~nm embedded in a polymer matrix with
      correlation length $\xi \sim 1$~nm. A repulsive interaction
      between the inclusion and the polymer produces a surface layer
      (with thickness $\sim \xi$) where the polymer segment density
      approaches zero. The local increase in free volume $v_f$ (within
      the {\em dash-dotted} circle) increases the penetrant
      diffusivity, leading to a significant increase in the bulk
      permeability when $\xi \ge a$ (see figure~\ref{fig:figure2}).}
  \end{center}
\end{figure}

Merkel {\em et al.}'s experiments exhibit {\em reverse selectivity},
meaning that the permeability of larger molecules is enhanced more
than smaller ones. This necessitates molecular-scale perturbations to
the polymer microstructure, precluding the notion that the
permeability is enhanced by a shell of void space between the
(impenetrable) inclusions and surrounding polymer. Rather, it suggests
a continuous increase in the free volume when approaching the
solid-polymer interface. Figure~\ref{fig:figure1} depicts a single
impenetrable sphere embedded in a polymer melt where polymer segments
are repelled from the solid surface. This is the microstructural view
that underlies the theory presented below.

Cohen and Turnbull's statistical mechanical theory~\citep{Cohen:1959}
yields a solute diffusion coefficient
\begin{equation} \label{eqn:candt}
  D = A \exp{(-\gamma v_m / v_f)}
\end{equation}
where $A$ and $\gamma$ are constants, $v_m$ is the minimum free volume
required for a solute molecule to escape its cage of neighboring atoms
and, hence, undergo diffusive migration, and $v_f$ is the available
free volume per volume occupying element. In this work, the volume
occupying elements are the atoms of the polymer melt. For simplicity,
each atom is assumed to occupy, on average, a volume $v_0$, where
$v_0^{1/3}$ is of the order of a covalent bond length ($\approx
1.5$~\AA). By considering the total volume, which comprises the sum of
free and occupied volume, it follows that
\begin{equation}  \label{eqn:e1}
  v_f = v_0 [m_1 / (v_0 n_1 \rho) - 1],
\end{equation}
where $c_0$ is the average atomic concentration, $m_1$ is the mass of
a monomer (repeat unit), $n_1$ is the number of atoms per monomer, and
$\rho$ is the (mass) density.

A tractable analytical expression for the radial polymer segment
density is obtained from a self-consistent mean-field model, with the
so-called ground-state approximation and a flat
interface~\citep{deGennes:1979}. The influence of curvature (finite
inclusion radius) is of secondary importance, and is therefore
neglected here. With a repulsive interaction between the polymer
segments and the solid, the segment concentration
is~\citep{deGennes:1979}
\begin{equation} \label{eqn:e2}
  c(r) = c_\infty \tanh^2[(r - a)/ \xi] + O(\xi / a)
\end{equation}
where $c_\infty$ is the bulk concentration, and $\xi = l / \sqrt{3 v
c_\infty}$ is the polymer correlation length, with $l$ the segment
length and $v$ the excluded volume (per segment). Conservation of
chain contour length and mass requires
\begin{equation}  \label{eqn:e3}
  \phi = c l^3 = (\rho / m_1) l_1 l^2,
\end{equation}
where $l_1$ is the length of a monomer. Combining
Eqns.~(\ref{eqn:candt})--(\ref{eqn:e3}) gives
\begin{equation} \label{eqn:candtc}
  D = D^\infty \exp{[- \frac{v_m^* \phi^* (\phi -
	\phi_\infty)}{(1-\phi^* \phi)(1-\phi^* \phi_\infty)}]},
\end{equation}
where $\phi^* = n_1 v_0 / (l_1 l^2)$ and $v_m^* = \gamma v_m / v_0$
are dimensionless parameters that reflect the prevailing atomic and
molecular geometry.

Let us simplify matters by assuming that the bulk polymer
concentration $\phi_\infty = c_\infty l^3 \approx 1$ and $v \approx
l^3$, so $l^2 \approx m_1 / (l_1 \rho)$, $\xi \approx l / \sqrt{3}$,
and $\phi^* \approx n_1 v_0 \rho / m_1$. Under these conditions, $1 -
\phi^*$ is the fractional free volume of the bulk polymer, and the
(maximum) diffusivity at the solid-polymer interface becomes
\begin{equation} \label{eqn:surface}
  D(r=a) \approx D^\infty \exp{[v_m^* \phi^* / (1 - \phi^*)]}.
\end{equation}
Clearly, the diffusivity at the interface can be much larger than in
the bulk when $\phi^* < 1$ and $v_m^* < 10$. Note that reverse
selectivity prevails, because the diffusivity decreases continuously
with increasing radial distance from the solid-polymer interface.

Consider, for example, bulk poly(p-trimethylsilyl styrene) (PTMSS),
for which $n_1 = 28$, $m_1 = 180.3$~g/mol, $l_1 \approx 3$~\AA \ (two
C-C bonds) and $\rho \approx
965$~kg/m$^3$~\cite{Khotimskii:2000}. With $\phi_\infty = 1$, $l
\approx 1.02$~nm, $\xi \approx 0.59$~nm and $\phi^* = 0.090 v_0$ (with
$v_0^{1/3}$ measured in \AA). To determine an appropriate value for
$v_0$, which, in principle, should not vary significantly from one
polymer to another, let us adopt a reported value of the fractional
free volume $1 - \phi^* = 0.191$~\cite{Khotimskii:2000}. This provides
$v_0^{1/3} \approx 2.0$~\AA \ and $v_f^{1/3} \approx 1.3$~\AA, which
are both of the expected magnitude.

With a statistically homogeneous microstructure, the average diffusive
flux can be expressed as a volume average
\begin{equation} \label{eqn:avgflux}
  \langle \vect{j} \rangle = V^{-1} \int_V \vect{j} \mbox{d}V,
\end{equation}
where the integration extends over the discrete and continuous phases
of a representative elementary volume $V$. The local diffusive flux is
$\vect{j} = - D \grad{n}$ (Fick's first law) where $n$ is the solute
concentration. Under steady conditions, conservation demands
\begin{equation} \label{eqn:cons}
  \dive{(D \grad{n})} = 0,
\end{equation}
where $D(r)$ is given by Eqns.~(\ref{eqn:e2})
and~(\ref{eqn:candtc}). When the volume fraction $\phi_p = n_p (4/3)
\pi a^3 \ll 1$, the average flux is~\citep{Hill:2005b}
\begin{equation} \label{eqn:avgfluxa}
  \langle \vect{j} \rangle \approx - D^\infty \langle \grad{n} \rangle
  + 3 \phi_p (B / a^3) D^\infty \langle \grad{n} \rangle +
  O(\phi_p^2),
\end{equation}
where $\langle \grad{n} \rangle = H \Delta p / L$ is the average
concentration gradient and $B$ is the dipole strength, \ie,
\begin{equation} \label{eqn:dipole}
  n \rightarrow \langle \grad{n} \rangle \cdot \vect{r} + B \langle
  \grad{n} \rangle \cdot \vect{r} r^{-3} \mbox{ as } r \rightarrow
  \infty.
\end{equation}
In this work, $B$ is obtained from an efficient numerical solution of
Eqn.~(\ref{eqn:cons}) that satisfies Eqn.~(\ref{eqn:dipole}) as $r
\rightarrow \infty$ with a no-flux boundary condition at $r =
a$\footnote{For impenetrable inclusions (\eg, non-porous silica), $D =
0$ when $r < a$.}.

The results are calculated with $\phi_\infty = 1$ and $\phi^* =
0.8$. The two remaining independent (dimensionless) parameters are the
scaled correlation length $\xi / a$ and the scaled tracer size $v_c^*
= \gamma v_m / v_0$. Here, the principal influence of $v_m^*$ is to
set the diffusion coefficient at the interface: $D(r=a) / D^\infty =
\exp{(4 v_m^*)}$.

As expected, all values of $v_m^* > 0$ increase the effective
diffusivity, because the available free volume increases with respect
to the unperturbed (bulk) polymer. Also, because the relative increase
in diffusivity depends exponentially on the penetrant size
(Eqn.~(\ref{eqn:surface})), situations with multiple diffusing species
exhibit reverse selectivity~\citep{Merkel:2002}.

\begin{figure}
  \begin{center}
    \vspace{1.0cm}
    \includegraphics[width=5cm]{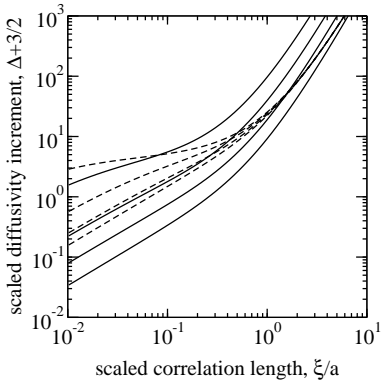}
    \vspace{1.0cm}
    \caption{\label{fig:figure2} The scaled diffusivity increment
      $\Delta = (D^e / D^\infty - 1) / \phi_p$ versus the scaled
      polymer correlation length $\xi / a$ with $v_m^* = \gamma v_m /
      v_0 = 0.1, $ 0.2, 0.4 and 1.0 (increasing upward), and $\phi^* =
      0.8$ ({\em solid lines}). The {\em dashed lines} are exact
      solutions (Eqn.~\ref{eqn:exact}) for impenetrable inclusions
      with a uniform coating of thickness $\xi$ where $D(a < r <a +
      \xi) / D^\infty = \exp{(4 v_m^*)}$.}
  \end{center}
\end{figure}

Figure~\ref{fig:figure2} presents the most general predictions of the
theory. The diffusivity increment, defined as
\begin{equation}
  \Delta = - 3 B / a^3 = (D^e / D ^\infty - 1)/ \phi_p + O(\phi_p),
\end{equation}
is plotted as a function of the scaled correlation length $\xi / a$
for various values of $v_m^*$ ({\em solid lines}). These calculations
are compared with the exact solution for an impenetrable sphere with a
uniform coating ({\em dashed lines}):
\begin{equation} \label{eqn:exact}
  \Delta = - 3 (1 + \xi / a)^3 \frac{(1 - \beta)[1 + \alpha (1 + \xi /
      a)^3] + 3 \beta}{(2 + \beta)[1 + \alpha (1 + \xi / a)^3] - 3
    \beta}.
\end{equation}
Here, $\beta = D / D^\infty$, with $D$ the diffusivity inside the
coating ($a < r < a + \xi$), $\xi$ is the coating thickness, and
$\alpha = 2$~\footnote{With $\alpha = (2 + D_0/D)/(1 - D_0/D)$,
Eqn.~(\ref{eqn:exact}) provides a solution for penetrable inclusions,
where $D_0$ is the diffusivity when $r < a$ and, again, $D$ is the
diffusivity when $a < r < a + \xi$.}. The figure demonstrates that the
effective diffusivity from Eqn.~(\ref{eqn:candt}) is much more
sensitive to the layer characteristics, particularly when the layers
are thick ($\xi / a > 1$).

\begin{figure}
  \begin{center}
    \vspace{1.0cm}
    \includegraphics[width=5cm]{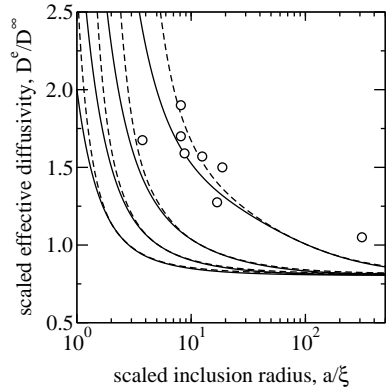}
    \vspace{1.0cm}
    \caption{\label{fig:figure3} The scaled effective diffusivity $D^e
      / D^\infty$ versus the scaled inclusion radius $a / \xi$ with
      $v_m^* = \gamma v_m / v_0 = 0.1, $ 0.2, 0.4 and 1.0 (increasing
      upward), $\phi^* = 0.8$, and $\phi_p = 0.13$: $O(\phi_p)$ (solid
      lines); $O(\phi_p^2)$ (dashed lines). The {\em circles} are
      experimental measurements of the permeability enhancement
      from~\cite{Merkel:2002} (reported radii scaled with $\xi =
      0.8$~nm).}
  \end{center}
\end{figure}

\begin{figure}
  \begin{center}
    \vspace{1.0cm}
    \includegraphics[width=5cm]{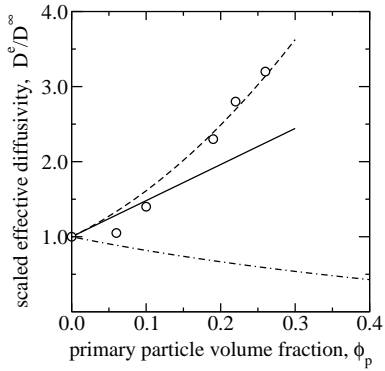}
    \vspace{1.0cm}
    \caption{\label{fig:figure4} The scaled effective diffusivity $D^e
      / D^\infty$ versus the inclusion volume fraction $\phi_p$ with
      $v_m^* = \gamma v_m / v_0 = 1.0$, $\phi^* = 0.8$ and $\xi / a =
      0.8 / 6.5 \approx 0.123$: exact $O(\phi_p)$ theory ({\em solid
	lines}); approximate $O(\phi_p^2)$ theory ({\em dashed
	lines}). The {\em circles} are experimental measurements of the
      permeability enhancement from~\cite{Merkel:2002} (with $a =
      6.5$~nm). The {\em dash-dotted line} is Maxwell's
      self-consistent theory for impenetrable inclusions and
      unperturbed (homogeneous) polymer.}
  \end{center}
\end{figure}

To compare the theory with the available experimental data,
figures~\ref{fig:figure3} and~\ref{fig:figure4} show the relative
effective diffusivity $D^e / D^\infty$ and measured values of the
relative permeability $D^e H / (D^\infty H^\infty)$ ({\em circles}),
with $\phi_p = 0.13$. The $O(\phi_p)$ theory ({\em solid lines})
neglects particle interactions, so the values are as given by
Eqn.~(\ref{eqn:avgfluxa}). The {\em dashed lines} are an $O(\phi^2)$
theory, which has elements of a self-consistent mean-field
approximation with an explicit correction for interactions between
pairs of particles in a statistically homogeneous
dispersion~\citep{Jeffrey:1973}. Note that the (single particle)
dipole strength and particle concentration both affect two-body
interactions.

Some of the experimental scatter in figure~\ref{fig:figure3} may be
attributed to the variety of filler particles (all embedded in
poly(4-methyl-2-pentyne)~(PMP)) and, possibly, different penetrant
species and degrees of particle
aggregation~\citep{Merkel:2002}. Nevertheless, with the foregoing
approximations, the correlation length inferred by the fit is $\xi =
0.8$~nm (with $\phi^* = 0.8$) and, hence, the segment length $l
\approx \sqrt{3} \xi \approx 1.4$~nm. The repeating unit of PMP
comprises two C-C bonds, so there are about 4.5 monomer units per
statistical segment, which is reasonable because polymers with much
less ``bulky'' side groups (\eg, poly(oxyethylene)) have fewer than
two repeat units per statistical
segment~\citep[\eg][pg. 168]{Russel:1989}.

Figure~\ref{fig:figure4} shows how the effective diffusivity (with $a
= 6.5$~nm) increases with the solid volume fraction. The theory is
presented with a correlation length $\xi = 0.8$~nm, which follows from
the data in figure~\ref{fig:figure3}, so $\xi / a = 0.8 / 6.5 \approx
0.123$. Despite the experimental data extrapolating to a value of $D^e
< D^\infty$ as $\phi_p \rightarrow 0$, the theoretical and
experimental trends are in good agreement. For reference, the {\em
dash-dotted} line is Maxwell's self-consistent theory for impenetrable
inclusions in an unperturbed polymer matrix. It is remarkable,
perhaps, that a perturbation extending only $\xi \approx 0.8$~nm from
the inclusion surfaces can have such a significant influence on the
bulk permeability.

Theoretical predictions of the effective diffusivity have been
compared with experimental measurements of the relative
permeability. Therefore, the solubility was assumed independent of the
inclusion concentration. Further, a mean-field description of the
polymer segment density was adopted, with a flat, repulsive interface
where the segment concentration vanishes (at $r = a$). Also, the
diffusion coefficient was assumed to follow Cohen and Turnbull's
formula (Eqn.~(\ref{eqn:candt})), with the occupied volume assumed
proportional to the atomic number density. Finally, theoretical
predictions were based on a statistically homogeneous dispersion of
inclusions, whereas the particles are often found to
aggregate. Nevertheless, the semi-quantitative agreement of theory and
experiment strongly suggests that reverse selectivity in these
nanocomposite membranes is of classical origin.

\begin{acknowledgments}
  Supported by the Natural Sciences and Engineering Research Council
  of Canada (NSERC), through grant number 204542, and the Canada
  Research Chairs program (Tier II). The author is grateful to
  M. Maric (McGill University) for helpful discussions related to this
  work.
\end{acknowledgments}

\bibliography{../../bibliographies/global.bib}
\bibliographystyle{plain}

\appendix

 \end{document}